\documentclass[twocolumn]{aastex62}

\submitjournal{ApJ}

\shorttitle{FMR from local variations}
\shortauthors{S\'anchez~Almeida \& S\'anchez-Menguiano}

\begin{document}

\title{The fundamental metallicity relation emerges from the local anti-correlation between star formation rate and gas-phase metallicity existing in disk galaxies}

\correspondingauthor{J. S\'anchez~Almeida}
\email{jos@iac.es}

\author[0000-0003-1123-6003]{J.\,S\'anchez~Almeida}
\affiliation{Instituto de Astrof\'\i sica de Canarias,\\
La Laguna, Tenerife, Spain}
\affiliation{Departamento de Astrof\'\i sica,\\
Universidad de La Laguna, Tenerife, Spain}

\author[0000-0003-1888-6578]{L. S\'anchez-Menguiano}
\affiliation{Instituto de Astrof\'\i sica de Canarias,\\
La Laguna, Tenerife, Spain}
\affiliation{Departamento de Astrof\'\i sica,\\
Universidad de La Laguna, Tenerife, Spain}




\begin{abstract}
The fundamental metallicity relation (FMR) states that galaxies of the same stellar mass but larger star formation rate (SFR) tend to have smaller gas-phase metallicity ($\langle Z_g\rangle$). It is thought to be fundamental because it naturally arises from the stochastic feeding of star-formation from external metal-poor gas accretion, a process extremely elusive to observe but essential according the cosmological simulations of galaxy formation. In this letter, we show how the FMR emerges from the local anti-correlation between SFR surface density and $Z_g$ recently observed to exist in disk galaxies. We analytically derive the global FMR from the local law, and then show that both relations agree quantitatively when considering the star-forming galaxies of the MaNGA survey. Thus, understanding the FMR becomes equivalent to understanding the origin of the anti-correlation between SFR and metallicity followed by the set of star-forming regions of any typical  galaxy. The correspondence between local and global laws is not specific of the FMR, so that a number of local relations should exist associated with known global relations.
\end{abstract}



\section{Rationale} \label{sec:intro}

The scaling relations followed by galaxies play a central role to assess our understanding of galaxy formation. The non-linear evolution of the density perturbations created during the Big Bang gives rise to galaxies in a sequence that can be followed only through numerical simulations. Many key physical processes (e.g., the feedback produced by stars and black holes) have to be taken into account through sub-grid recipes tuned to reproduce some of the scaling properties observed in galaxies (e.g., the distribution of stellar masses).  Other independent scaling relations allow us to test the consistency of the resulting model galaxies \citep[e.g.,][]{2014MNRAS.444.1518V,2015MNRAS.446..521S}.
Among them, the fundamental metallicity relation (FMR) stands out because it relates (and so constrains) the main ingredients involved in the star-formation process, its fueling, and its self-regulation. \citet{2008ApJ...672L.107E}, \citet{2010MNRAS.408.2115M}, and \citet{2010A&A...521L..53L} found out that galaxies of the same stellar mass with higher SFR show lower gas-phase metallicity. As it was already put forward by \citet{2008ApJ...672L.107E} and \citet{2010MNRAS.408.2115M}, the existence of this relation
suggests the  feeding of star-formation from external stochastic metal-poor gas accretion,
a process extremely elusive to observe \citep[e.g.,][]{2014A&ARv..22...71S} but essential according the numerical simulations of galaxy formation \citep[e.g.,][]{2009Natur.457..451D}. The accretion of metal poor gas triggers star-formation, which consumes the accreted gas, which eventually leads to reducing the SFR and to increasing the gas-phase metallicity. 
For recent reviews on the FMR, see, e.g., \citet{2017ASSL..430...67S}, \citet{2017MNRAS.469.2121S}, \citet{2018arXiv181106015C}, or \citet{2019A&ARv..27....3M}.

It has been recently found that star-forming disk galaxies are chemically inhomogeneous, with a decrease of metallicity spatially coinciding with enhanced  SFR \citep{2018MNRAS.476.4765S,2019ApJ...872..144H,2019arXiv1904.03930S}. Notably, this is not only a property of some rare extremely metal poor galaxies \citep[e.g.,][]{2013ApJ...767...74S,2015ApJ...810L..15S} but it happens to be characteristic of the full population of local star-forming galaxies with stellar mass $\log(M_\star/M_\odot)\lesssim 10.5$, as portrayed by MaNGA \citep{2019arXiv1904.03930S}. 

In this Letter, we show that the FMR emerges from the local anti-correlation between SFR surface density and metallicity.  We derive the global FMR from the local law (Sect.~\ref{sec:maths}), and then show that both relations agree observationally (Sect.~\ref{sec:observations}). The agreement is qualitative when considering FMRs from assorted references, but it becomes quantitative when using the same MaNGA galaxies for both the local and the global law  (Sect.~\ref{sec:observations}). Thus, understanding the FMR becomes equivalent to understanding the origin of the anti-correlation between SFR and metallicity followed by the set of star-forming regions of low-mass galaxies (Sect.~\ref{sec:conclusions}). The equivalence between local and global laws is to be expected also in other known global scaling relations (Sect.~\ref{sec:conclusions}).

%
\section{FMR from the local anti-correlation between $\Sigma_{\rm SFR}$ and $Z_{\MakeLowercase{g}}$}\label{sec:maths}

As we discussed in  Sect.~\ref{sec:intro}, the galaxies forming stars have been shown to present an anti-correlation between the SFR surface density, $\Sigma_{\rm SFR}$, and the gas-phase metallicity, $Z_g$, i.e.,
\begin{equation}
\Delta\log Z_g = m\,\Delta\log\Sigma_{\rm SFR},
\label{eq:approx1}
\end{equation}
with $\Delta X$ the residual of parameter $X$ once large scale variations have been removed or, equivalently, the difference between $X$ in two nearby regions of the same galaxy. Equation~(\ref{eq:approx1}) has been shown to hold by \citet{2019arXiv1904.03930S}, where the slope $m$, which quantifies the strength and sign of the correlation, is found to vary systematically depending on the global properties of the galaxy. Considering two galaxies with the same global properties, and thus the same $m$,  Eq.~(\ref{eq:approx1}) also describes the difference between two regions in the two galaxies, provided the regions are at the same galactocentric distance. Then the relation (\ref{eq:approx1}) also describes the excess of  $\log Z_g$ and $\log\Sigma_{\rm SFR}$ in one particular region of a particular galaxy with respect to the average alike galaxy\footnote{The set of {\em alike galaxies} to carry out the average includes all the galaxies with similar $m$ in Eq.~(\ref{eq:approx1})  whose integrated SFR and metallicity will be used to derive the FMR. It is a flexible definition because the implementation can mutate depending on the actual needs. In the example given in Sect.~\ref{sec:observations}, the {\em alike galaxies} are all the star-forming galaxies in MaNGA with similar stellar mass.} at the position of the region, i.e.,
\begin{equation}
\Delta\log Z_g = \log Z_g - \log Z_{g0},
\label{eq:approx99}
\end{equation}
\begin{displaymath}
\Delta\log \Sigma_{\rm SFR} = \log \Sigma_{\rm SFR} - \log \Sigma_{\rm SFR0},
\end{displaymath}
where $Z_{g0}$ and $\Sigma_{\rm SFR0}$ stand for the large scale variation of the gas-phase metallicity and SFR in the mean alike galaxy. Defining $\Delta Z_g = Z_g- Z_{g0}$ and $\Delta\Sigma_{\rm SFR} =\Sigma_{\rm SFR} -\Sigma_{\rm SFR0}$, and assuming $|\Delta Z_g| \ll Z_{g}$ and $|\Delta\Sigma_{\rm SFR}| \ll \Sigma_{\rm SFR}$, then Eq.~(\ref{eq:approx1}) can be re-written as 
\begin{equation}
\Delta Z_g/Z_{g0} \simeq m\,\Delta\Sigma_{\rm SFR}/\Sigma_{\rm SFR0}.
\label{eq:approx2}
\end{equation}

Note that, except for $m$, all variables in Eqs.~(\ref{eq:approx1}) and (\ref{eq:approx2}) depend on the position on the galaxy disk, defined by the Cartesian coordinates $x$ and $y$.  
Thus, the mean metallicity of the galaxy is defined to be 
\begin{equation}
\langle Z_g\rangle = \langle M_g\rangle^{-1}\,\int\,W\,Z_g\,\Sigma_g\,dxdy,
\label{eq:avmetallicity}
\end{equation}
\begin{displaymath}
\langle M_g\rangle = \int\,W\,\Sigma_g\,dxdy,
\end{displaymath}
with $\Sigma_g$ the surface density of the gas and $W$ an arbitrary weight that goes to zero outside the galaxy so that the integral extends to all the $x$-$y$ plane.  The symbol $\langle M_g\rangle$ stands for the total gas mass of the galaxy only if $W$ equals one within the galaxy. (To ease distinguishing local from galaxy-integrated quantities, the latter are always represented as symbols within brackets.) The excess mean metallicity can be defined as 
\begin{equation}
\Delta\langle Z_g\rangle = \langle Z_g\rangle - \langle Z_{g0}\rangle,
\label{eq:approx3}
\end{equation}
with $\langle Z_{g0}\rangle$ the mean metallicity to be expected considering all the alike galaxies. Equations~(\ref{eq:approx2}), (\ref{eq:avmetallicity}), and (\ref{eq:approx3}) lead to
\begin{equation}
\Delta\langle Z_g\rangle \simeq m\,\,\langle M_g\rangle^{-1}\,\int\,W\,Z_{g0}\,\Delta\Sigma_{\rm SFR}\frac{\Sigma_g}{\Sigma_{\rm SFR0}}\,dxdy.
\label{eq:master}
\end{equation}
The Kennicutt-Schmidt relation shows that the observed SFR surface density scales gas mass surface density through a law assumed to have the form of 
\begin{equation}
\Sigma_{\rm SFR} = A\,\Sigma_g^{N},
\label{eq:ksrelation}
\end{equation}
where $A$ and $N$ are numerical constants, with $N\simeq 1$ \citep[e.g.,][]{1998ApJ...498..541K,2008AJ....136.2846B}. Keeping only first order terms in both $N-1$ and $\Delta\Sigma_{\rm SFR}$,  Eq.~(\ref{eq:ksrelation}) becomes, 
\begin{equation}
\frac{\Sigma_g}{\Sigma_{\rm SFR0}}\simeq A^{N-2}\,\Big[1+(1-N)\,\ln\Sigma_{\rm SFR0}+\frac{\Delta\Sigma_{\rm SFR}}{\Sigma_{\rm SFR0}}\Big].
\label{eq:aksrelation}
\end{equation}
Introducing Eq.~(\ref{eq:aksrelation}) into Eq.~(\ref{eq:master}), and retaining only first order terms,
\begin{equation}
\Delta\langle Z_g\rangle \simeq m\,\langle{\rm SFR}\rangle^{-1}\int\,W\,Z_{g0}\,\Delta\Sigma_{\rm SFR}\,dxdy,
\label{eq:supermaster}
\end{equation}
with the integrated SFR defined as
\begin{equation}
\langle{\rm SFR}\rangle =\int\,W\,\Sigma_{\rm SFR}\,dxdy.
\label{eq:sfr}
\end{equation}
Equation~(\ref{eq:supermaster}) can be re-written as  
\begin{equation}
\Delta\langle Z_g\rangle \frac{\langle{\rm SFR}\rangle}{m} \simeq \langle Z_{g0}\rangle\,\big[\langle{\rm SFR}\rangle-\langle{\rm SFR0}\rangle\big] +
\label{eq:approx10}
\end{equation}
\begin{displaymath}
\int\,W\,\big(Z_{g0}-\langle Z_{g0}\rangle\big)\,\Delta\Sigma_{\rm SFR}\,dxdy.
\end{displaymath}
The second term in the right-hand side of Eq.~(\ref{eq:approx10}) is negligible if the local excess of SFR ($\Delta\Sigma_{\rm SFR}$) is uncorrelated with the large scale variations of metallicity ($Z_{g0}-\langle Z_{g0}\rangle$).   In principle, local and large scale variations are expected to be independent from each other, allowing the condition to be met.  However, for the moment, we neglect this term just as an ansatz to be justified later on in  Sect.~\ref{sec:observations}  using real galaxies. Thus, one can rewrite Eq.~(\ref{eq:approx10}) as 
\begin{equation}
\Delta\langle Z_g\rangle/\langle Z_{g0}\rangle \simeq m\,\Delta\langle{\rm SFR}\rangle/\langle{\rm SFR}\rangle,
\label{eq:global}
\end{equation}
or, neglecting 2nd order terms,
\begin{equation}
\Delta\log\langle Z_g\rangle \simeq m\,\Delta\log\langle{\rm SFR}\rangle.
\label{eq:globalog}
\end{equation}

Equation~(\ref{eq:globalog}) shows the global FMR, which emerges from the local anti-correlation given in Eq.~(\ref{eq:approx1}) with exactly the same slope. 

\section{Comparison with observations}\label{sec:observations}

\begin{figure}[ht!]
\centering
\includegraphics[width=1.0\linewidth]{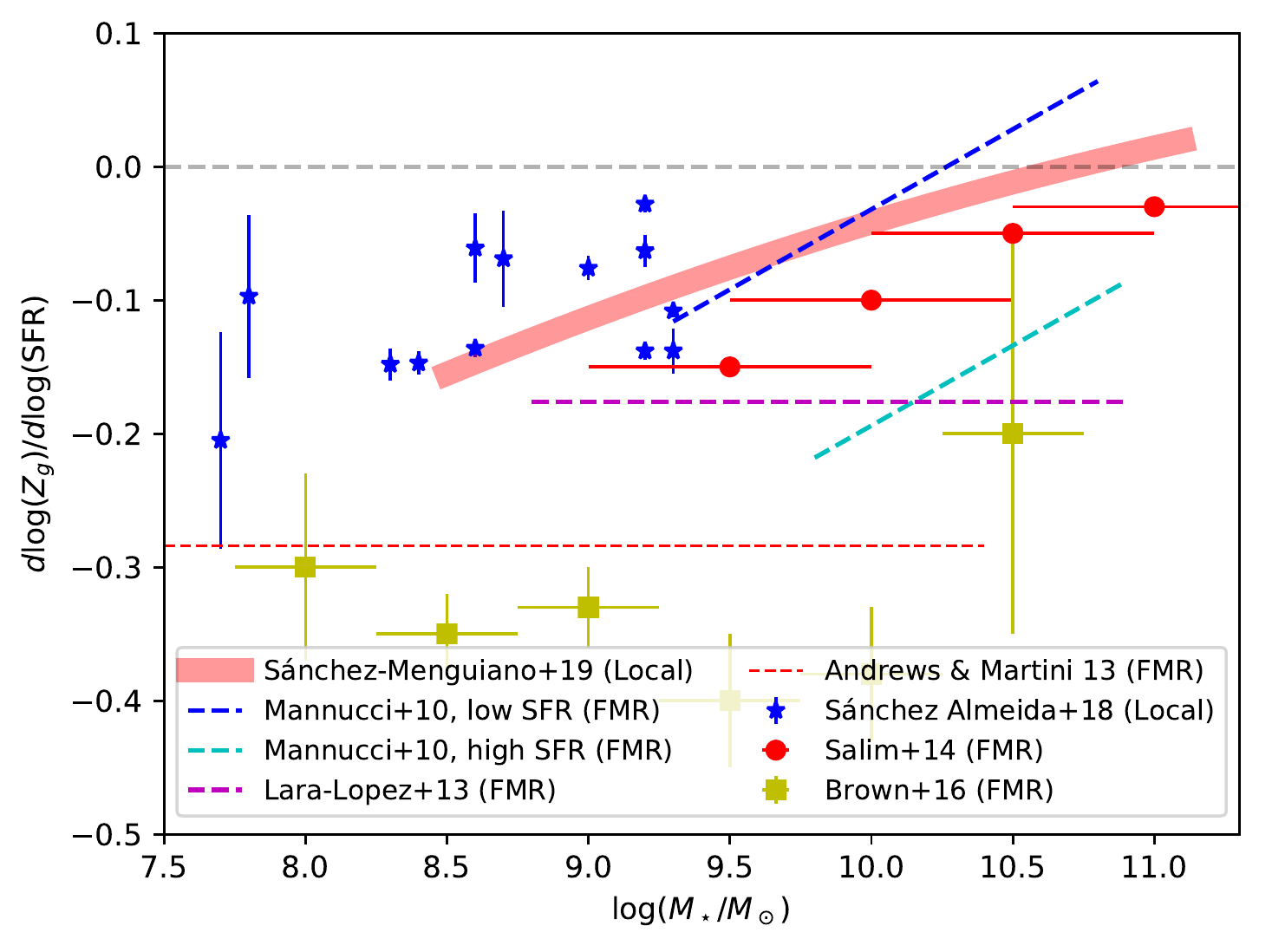}
\caption{Slope $m$ versus $M_\star$. The references cited in the inset are 
\citet{2019arXiv1904.03930S},
\citet{2018MNRAS.476.4765S},
\citet{2010MNRAS.408.2115M},
\citet{2013ApJ...764..178L},
\citet{2014ApJ...797..126S},
\citet{2016MNRAS.458.1529B},
\citet{2013ApJ...765..140A}, with the tag {\em Local} or {\em FMR} indicating whether $m$ was inferred from local variations or from a FMR.
The gray dashed line corresponds to $m=0$.
 \label{fig:ref}}
\end{figure}

\citet{2019arXiv1904.03930S} measured the slope $m$ in a set of 736 nearby spiral galaxies from the DR15 SDSS IV MaNGA survey \citep{2015ApJ...798....7B}. Galaxies were selected  the full MaNGA parent sample to be both star-forming and large enough to allow measuring local variations of SFR and $Z_g$. The value of $m$ was inferred from a linear fit to the scatter plot $\Delta Z_g$ versus $\Delta\Sigma_{\rm SFR}$ in each individual galaxy computed after removing the large scale trends. $Z_g$ and SFR were deduced using the strong line ratio O3N2 and the H$\alpha$ flux, respectively \citep[for details, see][]{2019arXiv1904.03930S}.  The slope $m$ was found to depend on $M_\star$, going from negative at low mass ($\log[M_\star/M_\odot]\lesssim 10.5$) to slightly positive at the high mass end of the mass distribution ($\log[M_\star/M_\odot]\simeq 11$). The thick red line in Fig.~\ref{fig:ref} shows the resulting mean $m$ for a given $M_\star$.
Figure~\ref{fig:ref} also includes the slopes worked out by \citet[][ blue stars with error bars]{2018MNRAS.476.4765S} for to the local anti-correlation found in a number of dwarf galaxies of the local Universe, which agree quite well with the estimates by \citeauthor{2019arXiv1904.03930S}
Slopes inferred from other assorted FMRs in the literature are also included in Fig.~\ref{fig:ref}  (references are given in the inset).  The slopes corresponding to the FMRs were computed from the published equation $Z_g=Z_g({\rm SFR},M_\star)$ as 
\begin{equation}
m = \frac{\partial\log Z_g}{\partial\log{\rm SFR}}\Big|_{M_\star}.
\end{equation}
As one can see from Fig.~\ref{fig:ref}, the $m$s inferred from local and global variations agree qualitatively; the slopes are negative, and there is a tendency for $|m|$ to decrease as $M_\star$ increases \citep[see, e.g., the values from ][]{2014ApJ...797..126S}. Qualitative differences are also evident, though. We think that these differences are mostly due to systematic errors in the various calibrations used to estimate $Z_g$ and SFR \citep[e.g.,][]{2008ApJ...681.1183K,2012ARA&A..50..531K}, and also due to the actual set of galaxies employed to infer $m$ \citep[e.g.,][]{2014ApJ...797..126S}. This impression is corroborated by the exercise described in the next paragraph, where local and global laws obtained using the same galaxies and identical $Z_g$ and $\Sigma_{\rm SFR}$ are shown to coincide.    

\begin{figure*}[ht!]
\centering
\includegraphics[width=0.9\linewidth]{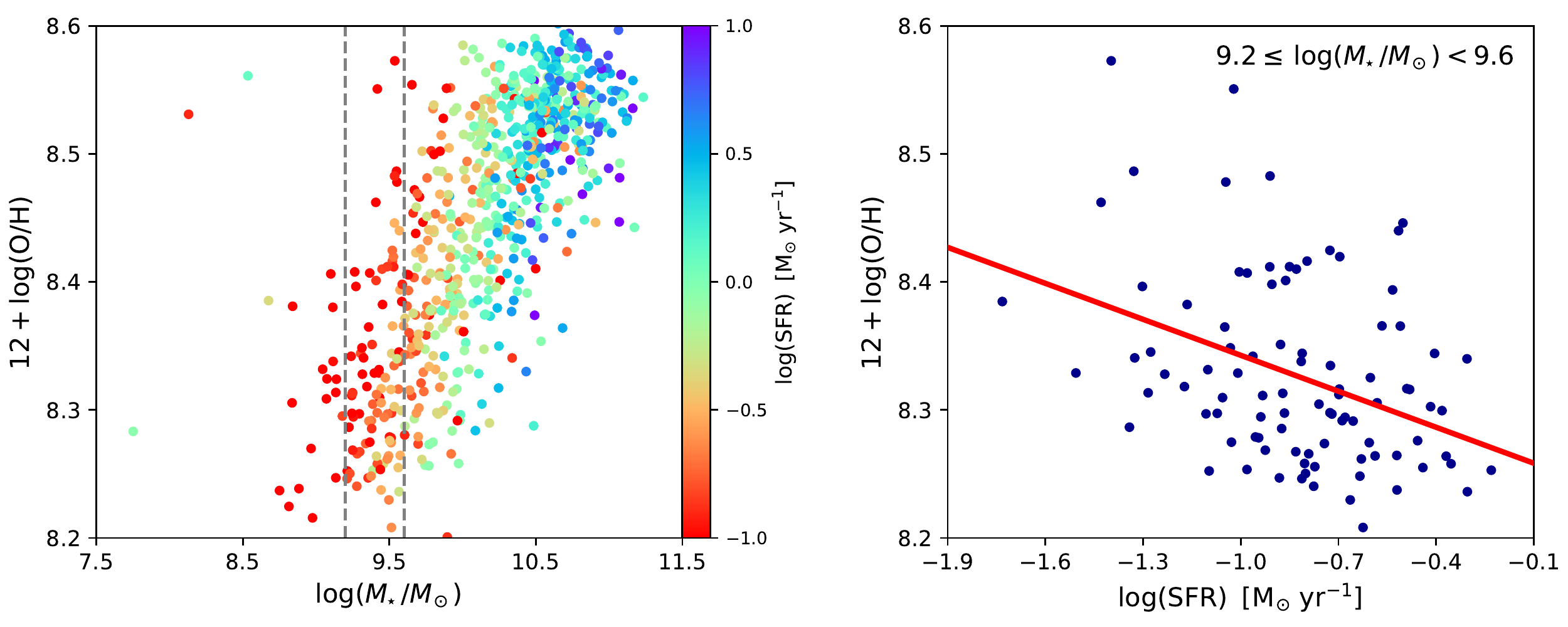}
\caption{Left panel: FMR for the 736 MaNGA galaxies used by \citet{2019arXiv1904.03930S}, i.e., $\langle Z_g\rangle$ versus $M_\star$  color-coded with SFR. Given $M_\star$, galaxies of lower  $\langle Z_g\rangle$ tend to have higher SFR. Right panel: Scatter plot $\langle Z_g\rangle$ versus SFR for the galaxies in the mass bin $9.2\leq \log(M_\star/M_\odot) \leq 9.6$, which is marked with dashed lines in the left panel. The slope of the linear fit shown in red renders $m$. \label{fig:procedure}}
\end{figure*}
\begin{figure}[ht!]
\includegraphics[width=1.0\linewidth]{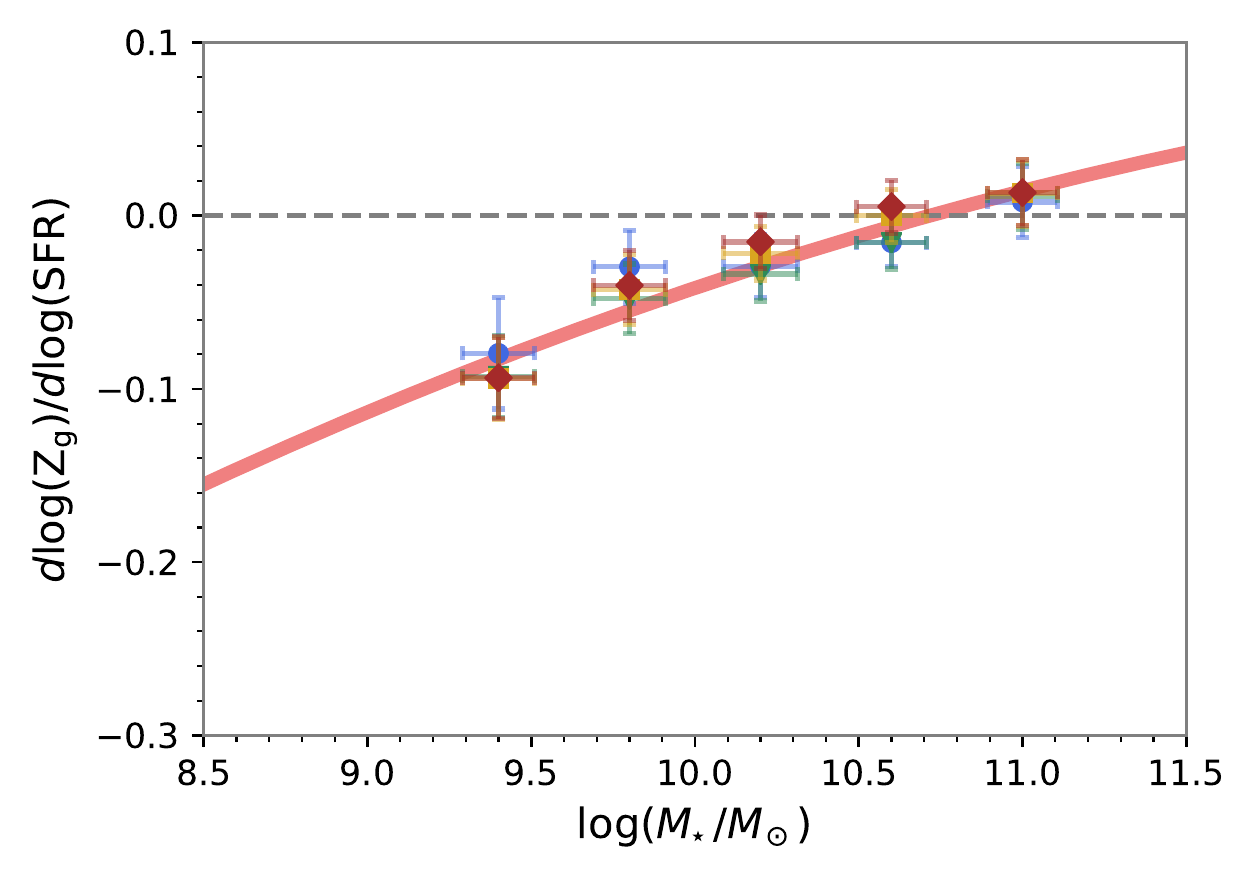}
\caption{Comparison of the slope $m$ inferred from local and global FMR relations derived from  exactly the same data set. The solid line represents slopes obtained from spatially resolved MaNGA galaxies \citep{2019arXiv1904.03930S}. The symbols represent slopes obtained from integrated values of metallicity and SFR using exactly the same galaxies. Different colors represent different ways of computing the average metallicity (see text for details). The error bars correspond to the uncertainty associated with the determination of the slope (vertical), and the RMS variation of the stellar masses of the galaxies in the mass bin used to infer $m$ (horizontal). \label{fig:ref_manga}}
\end{figure}

 
The equivalence between the local properties and the integrated quantities is a mathematical identity, therefore,  it holds provided certain conditions are met.  The main ones are: (1) perturbations are small so that second and higher order terms can be neglected and (2)  large-scale variations of metallicity and small-scale variations of SFR are uncorrelated. In order to check whether these two conditions are usually met, we have computed $\Delta$SFR and $\Delta\langle Z_g\rangle$ for the same 736 MaNGA galaxies from which \citet{2019arXiv1904.03930S} worked out the {\em local} anti-correlation between $\Delta\Sigma_{\rm SFR}$ and $\Delta Z_g$ shown in  Fig~\ref{fig:ref}. We aim at comparing the slope inferred from integrated quantities, without approximations (Eqs.~[\ref{eq:avmetallicity}], [\ref{eq:sfr}], and [\ref{eq:globalog}]), with those obtained from the spatially resolved quantities (Eq.~[\ref{eq:approx1}]). Mean metallicities are derived in four different ways to test the dependence of the conclusions on the way the spatial average $\langle Z_g\rangle$ is computed. The result for weighting with $\Sigma_{\rm SFR}$  is shown in Fig.~\ref{fig:procedure}\footnote{$Z_g$ is the fraction of metals by mass, whereas O/H in Fig.~\ref{fig:procedure} is the abundance of O relative to H by number. There is no inconsistency in using  any of them to infer $m$ since O/H and $Z_g$ are proportional to each other, provided the relative abundance of the different metals is the same for all galaxies.} (according to Eq.~[\ref{eq:ksrelation}], $\Sigma_{\rm SFR}$ is a proxy for $\Sigma_{g}$ when $N=1$). We employ Eq.~(\ref{eq:avmetallicity}) replacing $\Sigma_g$ with $\Sigma_{\rm SFR}$, and setting $W$ to a constant $\not= 0$ only in the star-forming regions. MaNGA galaxies follow the global FMR as evidenced in Fig.~\ref{fig:procedure}, left panel, where the mass metallicity relation is color-coded with SFR. It is also clear from Fig.~\ref{fig:procedure}, right panel, showing the scatter plot  $\langle Z_g\rangle$ versus SFR for all galaxies within a narrow range of stellar masses ($9.2 \le \log[M_\star/M_\odot] \leq 9.6$). $Z_g$ and $\Sigma_{\rm SFR}$ were computed with exactly the same recipes employed by \citet{2019arXiv1904.03930S}, and the actual galaxies and the star-forming regions within them were also the same.   
A linear fit to scatter plots like the one in  Fig.~\ref{fig:procedure}, right panel, provides $m$ at a particular mass range. Repeating the exercise for different mass bins renders the dependence of $m$ on $M_\star$ represented in Fig.~\ref{fig:ref_manga}.\footnote{As discussed by, e.g., \citet{2017MNRAS.469.2121S,2019MNRAS.484.3042S}, the use of finite mass bins may induce a spurious correlation between $\langle Z_g\rangle$ and SFR. However, this artificial correlation is always positive since both $\langle Z_g\rangle$ and SFR tend to increase with increasing $M_\star$. Thus, $m \leq 0$ cannot result from such a bias. As a sanity check,  we verified that dividing by half the mass bin does not produce a significant variation in Fig.~\ref{fig:ref_manga}.} In order to test de  dependence of the result on the way the average $\langle Z_g\rangle$ is carried out,   Fig.~\ref{fig:ref_manga} shows the result based on weighting with $\Sigma_{\rm SFR}$ (the red symbols), with $\Sigma_{\rm SFR}^{0.7}$(the orange symbols), without weighting  (the blue symbols), and using the metallicity at the effective radius (the green symbols). The three first averages correspond to weighting with $\Sigma_g$ if the Kennicutt-Schmidt relation holds (Eq.~[\ref{eq:ksrelation}]) and $N=1$, 1.4, and 0, respectively. $N=1.4$ corresponds to the canonical value in the calibration by \citet{1998ApJ...498..541K}.

The three main conclusions ensuing from Fig.~\ref{fig:ref_manga} are: (1) there is a very good agreement between the slope derived from local and global laws, which reinforces the approximations leading to the analytical derivation in Sect.~\ref{sec:maths}. (2) The way in which the average is computed does no seem to be very important. Four different ways of averaging yield similar results. (3) The number of galaxies used to construct Fig.~\ref{fig:ref_manga} is large enough for the conclusions to be representative of local galaxies (as portrayed by MaNGA). 

%
\section{Discussion and conclusions}\label{sec:conclusions}
Provided the existence of a local correlation between $\Sigma_{\rm SFR}$ and $Z_g$, the FMR is a must -- Eq.~(\ref{eq:approx1}) $\Rightarrow$ Eq.~(\ref{eq:globalog}). Does the inverse hold true, i.e., does the existence of the FMR imply the existence a local anti-correlation? The answer is {\em yes}. If a galaxy does not have a local correlation, it cannot show a correlation in integrated quantities (i.e., if $m=0$ in Eq.~[\ref{eq:approx1}], $m$ cannot be $\not= 0$ in Eq.~[\ref{eq:globalog}]). Thus, understanding the FMR becomes equivalent to understanding the origin of the correlation between SFR and metallicity followed by the set of star-forming regions of any typical galaxy. 

A potential flaw of the above argument is the break down of the approximations used to derive Eq.~(\ref{eq:globalog}) from Eq.~(\ref{eq:approx1}), namely, the small variation of the physical parameters, and the lack of correlation between large-scale metallicity variations and small-scale SFR variations. The validity of these approximations was tested using a set of 736 star-forming galaxies in MaNGA, which give the same slope $m$ from local variations and from spatially integrated quantities (Fig.~\ref{fig:ref_manga}). We note that this set is representative of the local star-forming galaxies with masses in the range $8.7\leq \log(M_\star/M_\odot)\leq  11.2$. Still, tt may break down when considering other more extreme galaxies observed at higher spatial resolution.

The exercise with MaNGA galaxies also allows us to conclude that the actual weight employed to compute the integrated quantities is not fundamental. On the other hand, the methods to estimate SFR and $Z_g$ from emission line fluxes, and the set of galaxies included in the analysis, do influence the actual value inferred for $m$ (Fig.~\ref{fig:ref}).   

Our derivation in Sect.~\ref{sec:maths} is quite general, and shows that any local relation between two variables gives rise to a global relation. This is also true for the gas mass, for which an {\em alternative} FMR is known to exist. Galaxies with larger gas mass (both atomic and molecular) have smaller $Z_g$ \citep[e.g.,][]{2016A&A...595A..48B,2016MNRAS.455.1156B}. The fact that this global law has been observed implies that there should be a local anti-correlation between $\Sigma_g$ and $Z_g$. Indeed, \citet{2018ApJ...852...74B} find a local anti-correlation between the {\em gas fraction} inferred from dust extinction and $Z_g$. Similarly, \citet{2015MNRAS.446.1449L}  analyze a set of local analogues to Ly break galaxies, finding that objects with younger stellar populations have lower metallicity at a fixed mass. This would imply the existence of a local relation between the age of the stellar population and metallicity of the gas forming stars. These specific predictions of our equations can be tested observationally.

\acknowledgments
We are indebted to Sebasti\'an S\'anchez for  enlightening  discussions which  helped us to sharpen the arguments in the Letter, and to an anonymous referee for suggestions to clarify the presentation of the equations in Sect.~\ref{sec:maths}. 
The work has been partly funded by the Spanish Ministry of Economy and Competitiveness (MINECO), project AYA2016-79724-C4-2-P (ESTALLIDOS).
We use the MaNGA-Pipe3D data products created by the IA-UNAM MaNGA team under ConaCyt-180125 project.
Funding for the Sloan Digital Sky Survey IV has been provided by the Alfred P. Sloan Foundation, the U.S. Department of Energy Office of Science, and the Participating Institutions.The SDSS web site is www.sdss.org.
%

\vspace{5mm}
\facilities{MaNGA \citep{2015ApJ...798....7B}}


\software{astropy \citep{2013A&A...558A..33A}
          }




\end{document}